\documentclass{moriond}

\def\be{\begin{equation}}
\def\ee{\end{equation}}
\def\bea{\begin{eqnarray}}
\def\eea{\end{eqnarray}}

\usepackage{courier}
\usepackage{tikz}
\usetikzlibrary{shapes, snakes}
\usetikzlibrary{calc}
\usetikzlibrary{decorations.pathreplacing}
\usetikzlibrary{intersections}
\definecolor{cadmiumgreen}{rgb}{0.0, 0.42, 0.24}
\definecolor{bronze}{rgb}{0.8, 0.5, 0.2}
\newcommand{\Photo}{}

\begin{document}
\vspace*{4cm}
\title{Applying TDI to Hardware-Simulated Data}

\author{ Reid Ferguson$^1$, Olaf Hartwig$^1$, and Guido Mueller$^{1,2,3}$ }

\address{$^1$Max Planck Institute for Gravitational Physics, Appelstrasse 9, D-30167 Hanover, Germany \\ 
    $^2$University of Florida, P.O. Box 118440, Gainesville, FL 32611-8440, USA\\
    $^3$Leibniz University Hanover, Callinstr. 38, D-30167 Hanover, Germany}

\maketitle\abstracts{
    Many years of development have gone into producing instruments that meet the required noise performance of the LISA interferometric detection system. Concurrently, software simulations have been used to extensively develop the data analysis libraries to be used in the LISA pipeline, not least among which are the Time Delay Interferometry (TDI) algorithms. To bridge the gap between these two, we are developing a hardware-in-the-loop testbed to apply realistic, time-varying delays to signals traveling between phasemeters. We have shown that the testbed adds a sufficiently low amount of noise across the entire relevant LISA spectrum. We have also injected a realistic gravitational wave signal, generated via the LISA Data Challenge codebase, and successfully extracted it using TDI to remove the obscuring frequency noise of the carrier signal. Future efforts will expand the testbed to create a representative simulation of the entire LISA constellation, with an eye towards its use as a tool to aid in the development of the LISA data analysis pipelines. }

\section{Time Delay Interferometry in LISA}

% The LISA mission will use a constellation of three spacecraft to measure gravitational waves. The spacecraft will be separated by 2.5 million kilometers, and the laser beams sent between them will be delayed by this distance. This delay can be compensated for using time delay interferometry (TDI), allowing us to extract the gravitational wave signal from beneath the noise in the 280 THz laser carrier frequency, which couples into the readout the same way as the gravitational wave strain measurement. The TDI algorithms are designed to remove the laser frequency noise, and the resulting signal is then passed to the data analysis pipeline for further processing.

In terrestrial gravitational wave observatories, the near-equal arm lengths of the interferometers and the ability to stabilize the lasers to the common arm cavities decouple laser frequency noise from the readout of the GW signal. In the case of LISA, the arm lengths are not only unequal, but also time-varying by a few percent over the orbital period, and arm locking is neither foreseen nor sufficient in accomplishing the same in space.\cite{Ghosh_2022} The TDI algorithms are designed to take advantage of the fact that the laser frequency noise is common to all three spacecraft, and can be removed by taking linear combinations in post-processing of the signals from each spacecraft.

% The principle of these algorithms is simple: each laser is measured locally before it is transmitted across to the distant spacecraft and measured there. The same noise exists independently in both of these measurements. Shifting one measurement in time until the noise cancels out is the basis of TDI --- in essence, an equal-arm-length Michelson interferometer constructed in post-processing from linear combinations of the time series of each measurement. 
% \subsection{Transponder Signals}
% The delay in a signal traveling between spacecraft is formalized in the \emph{delay operator} $\mathbf{D}_{ij}(t)$. If $\phi_1$ is the local phase of the laser on Spacecraft 1 (S/C 1), then 
% \begin{equation}
%     \mathbf{D}_{12}(t)\phi_2 = \phi_2(t - d_{12}(t))
% \end{equation}
%  is the phase of the received field coming from S/C 2, and $d_{12}(t)$ is the proper time along a null geodesic from S/C 2 to S/C 1 at time $t$. 

% In the context of constructing the TDI combinations, the physical delay of a signal is represented by $\mathbf{D}_{ij}$, whereas the timeshifts applied in post-processing are represented by $\mathcal{D}_{ij}$. Multiple nested delay operators combine as 

% \begin{equation}
%     \mathcal{D}_{ij}\mathcal{D}_{jk}...\mathcal{D}_{nm} = \mathcal{D}_{ijk...nm}
% \end{equation}

The core element of the TDI combinations is the single-link inter-spacecraft signal $\eta_{ij}$, a MHz beatnote at frequency $\Delta \omega_{ij}$ which forms at the primary beamsplitter of the inter-spacecraft interferometer (ISI):

\begin{equation}
    \label{eq:eta}
    \eta_{ij} = \Delta \omega_{ij} t + \phi_i - \mathbf{D}_{ij}\phi_j
\end{equation}
where $\mathbf{D}_{ij}$ encodes the travel time of the signal sent by Spacecraft $j$ and received by Spacecraft $i$. These beatnotes are slowly varying between 5 and 25 MHz due to changing Doppler shifts induced by relative motion between the satellites.

The task of producing a hardware simulation system therefore reduces to producing a realistic recreation of these inter-spacecraft measurements --- a MHz beatnote between a delayed and an undelayed signal. The delayed signal must be able to be phase- and frequency-modulated to simulate the effects of orbital mechanics and gravitational wave strain. 

The TDI combinations are then formed by taking linear combinations of these signals, which can be expressed in terms of the delay operators. For the demonstration provided here, we simulated a static constellation with unequal arms. This corresponds to the TDI X1.0 combination,\cite{hartwig2021instrumental} which is given by

\begin{equation}
    X_1 = (1 - \mathcal{D}_{121})(\eta_{13} + \mathcal{D}_{13}\eta_{31}) - (1 - \mathcal{D}_{131})(\eta_{12} + \mathcal{D}_{12}\eta_{21}), 
    \label{tdieq}
\end{equation}

with $\mathcal{D}$ representing the time shifts applied in post-processing, contrasting with $\mathbf{D}$ representing physical delays of a signal (as in Eq. \ref{eq:eta}). Successive subscripts represent nesting, i.e. $\mathcal{D}_{121} = \mathcal{D}_{12}\mathcal{D}_{21}$.

% \subsection{Clocking Sidebands}

% To implement the correct delay operators in post-processing, the sampling times of each signal must be synchronized. The rate at which the clocks on board the spacecraft count time is slightly different --- and sufficiently enough so as to ruin the sensitivity of the observatory. To correct for this, each device sampling a transponder signal also generates a 2.4 or 2.401 GHz tone which is phase-modulated onto the local carrier before it goes into the combining beamsplitter and/or is sent to the distant spacecraft. 

% The combined single-link signal therefore also contains sidebands coming from the interference between these clock tone sidebands, which is then interfered with the carrier-carrier beatnote. It suffices to describe the relevant combination as the $5 .. 25$MHz carrier-carrier beatnote previously mentioned, and a relative clocking beatnote that is 1 MHz away from the carrier beatnote. The latter is used to correct for timing jitter, while the former is used to produce the TDI combinations and whose phase evolution encodes the strain perturbations along its path. More information can be found in §5.3.3 of the LISA Definition Study Report.\cite{colpi2024lisadefinitionstudyreport} Pseudo-random noise codes for ranging and data transfer are also modulated onto the carriers, but this will not be discussed in these proceedings.

\section{Hardware Simulation}
 The hardware testbed is designed to be a low-noise, low-latency system that can be used to test the TDI algorithms and the data analysis pipeline. As shown in Figure 1, the testbed consists of a series of hardware components that simulate the LISA spacecraft and the laser fields exchanged between them. The setup includes a signal generator that is currently only used to generate a monochromatic tone that mimics the carrier laser light from one spacecraft. This is fed to a Xilinx ZCU208 that digitizes the signal and sends it via an FPGA to RAM for a time-variable delay to simulate inter-spacecraft distance. Interfacing with the FPGA via the on-chip processing system (PS) is done to modulate the delay times, add Doppler shifts, and inject arbitrary gravitational waves into the delayed carrier.

% The primary device used is a ZCU208 evaluation board by AMD Xilinx, which sports an UltraScale+ RFSoC FPGA and 16GB of RAM --- 8GB for the programmable logic (PL) and 8GB for the processing system (PS)--- as well as 8 ADC and 8 DAC channels capable of sampling up to 5GHz \footnote{This is too high for the RAM to handle; it is set to 2.048GHz, which is then downsampled to 128MHz}. The FPGA, and the delay line architecture programmed into it, are controlled via a Jupyter server running on a lightweight PetaLinux operating system on the PS, and is accessible via Ethernet. 

\subsection{Delay Line Performance}\label{subsec:acf}

 It can be verified experimentally that the noise added by the delay is due to timing jitter, or clock noise, differences between the FPGA board and external devices. To be more specific, the signal at frequency $\omega$ with phase $\phi$ generated by a Moku:Pro phasemeter picks up phase jitter $+\omega\delta t_M$.\footnote{These jitters, $\delta t$, couple in with a minus sign when a signal is \emph{digitized} by an ADC and with a plus sign when it is \emph{synthesized} by a DAC.} It is then sampled by the delay line board, picking up $-\omega\delta t_D(t)$. If it is immediately sent back out of the delay line's DAC and measured, these cancel out to first order. Otherwise, upon re-synthesis it picks up $+\omega\delta t_D(t)$, but still contains the phase jitter $-\omega\delta t_D(t -\tau)$. When digitized by the Moku phasemeter, it finally picks up $-\omega\delta t_M(t - \tau)$. The phase of the measured signal with frequency $\omega$ is therefore given by

\begin{equation}
    \begin{array}{rc@{\,}c@{\,}l}
    \phi_\tau(t) &= &&\phi(t - \tau) + \omega\left(\delta t_M(t - \tau) - \delta t_D(t - \tau) + \delta t_D(t) - \delta t_M(t)\right) \\\\
    &= &&\mathbf{D}_\tau\phi(t) + \mathbf{D}_\tau\omega(t)\left(1 - \mathbf{D}_\tau\right)\left(\delta t_D(t) - \delta t_M(t)\right).
    \end{array}
    \label{eq:clocktrace}
\end{equation}

The lattermost term, the differential timing jitter, can be measured by generating a tone on the delay board via Direct Digital Synthesis (DDS) and measuring it, whereby it picks up the relevant jitters. 
\begin{equation}
    \begin{array}{rc@{\,}c@{\,}l}
    \phi_{DDS} &= &&\omega_{DDS}(t + \delta t_D(t) - \delta t_M(t)) = \phi_{0,DDS} + \tilde{\phi}_{DDS}\\\\
    \rightarrow \phi_\tau(t) &=&& \mathbf{D}_\tau\phi(t) + \frac{\omega}{\omega_{DDS}}\left(1 - \mathbf{D}_\tau\right)\tilde\phi_{DDS}(t).
    \end{array}
\end{equation}

A diagram of this measurement is shown in Figure 1, with the results plotted below it. The spectral density of the signal $S$ is then given by 
\begin{equation}
    S[\phi_\tau](f) =  S[\phi](f) + \left(\frac{\omega}{\omega_{DDS}}\right)^24\sin^2(\pi f\tau) S[\phi_{DDS}](f),
    \label{eq:cjmodelASD}
\end{equation}

where the sine term is sourced by the Fourier transform of the delay operator, $e^{-i2\pi f\tau}$.
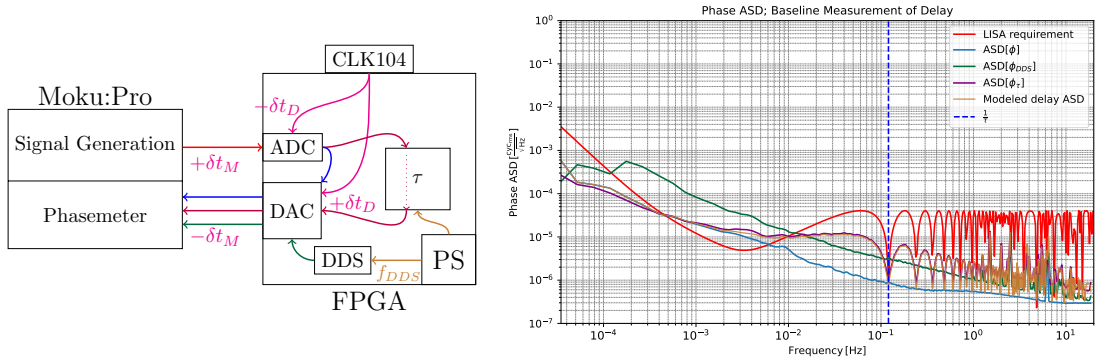
\begin{figure}
        \label{fig:clockjitter}
    \centering
    \begin{minipage}{0.4\linewidth}

        \scalebox{0.7}{
        \begin{tikzpicture}
            \node (moku) at (0,0) [draw, rectangle split, rectangle split parts=2, inner ysep = 0.5cm] {Signal Generation \nodepart{two} Phasemeter};
            \node (mokulabel) at ($(moku.north)+(0,0.25)$) [] {\Large Moku:Pro};
            \node (fpga) at ($(moku.east) + (1.5cm,0)$) [draw, rectangle, inner xsep=2cm, inner ysep=2cm, anchor=west] {};
            \node (fpgalabel) at ($(fpga.south)+(0,0)$) [anchor=north] {\Large FPGA};
            \node (clk104) at (fpga.north) [draw, rectangle, anchor=south] {CLK104};
            \node (sgpin) at ($(moku.east) + (0,0.6cm)$) []{};
            % \node (sglabel) at (sgpin) [above right]{\color{blue} A};
            \node (epsm) at (sgpin) [below right]{\color{magenta} $+\delta t_M$};
            \node (adc) at (fpga.west |- sgpin) [draw, rectangle, anchor=west] {ADC}; 
            % \node (adclabel) at (adc.east) [above right]{\color{blue} B};
            \node (del) at (fpga) [draw, rectangle, inner xsep=0.5cm, inner ysep = 0.5cm, anchor=west, xshift=0.3cm] {$\tau$};
           
            \node (pm) at ($(moku.east)-(0,0.6)$)[anchor=east]{};
            % \node (pmlabel) at (pm.east) [left, yshift = 0.1cm]{\color{blue} D};
            \node (dac) at (fpga.west |- pm) [draw, rectangle, anchor=west, inner ysep=0.4cm] {DAC};
            % \node (daclabel) at (dac.east) [above right]{\color{blue} C};
            \draw[->, thick, magenta] (clk104) to [out=270, in=0] ($(dac.north east) - (0,0.2)$) node[below right, yshift=0.1cm] {$+\delta t_D$};
           
            \draw[->, thick, red] (moku.east |- sgpin)--(adc.west);
            \draw[->, thick, purple] (adc.east) to[out=0, in=110] (del);
                \node (arrowstart) at (del.110) [] {};
                \node (arrowend) at (arrowstart|-del.south) [anchor=north] {};
            \draw[->, dotted, purple] (arrowstart)--(arrowend);
            \draw[->, thick, purple] (arrowend.north) to[out=270, in=0] (dac);
            \draw[->, thick, blue] (adc) to[out=0, in=north east] (dac);
            \draw[->, thick, blue] (dac.155)--(dac.155 -| pm.east);
            \draw[->, thick, purple] (dac.west)--(pm.east);
            \draw[->, thick, cadmiumgreen] (dac.205)--(dac.205 -| pm.east);
            \node (epsm) at (pm.east) [below right, yshift=-0.2cm]{\color{magenta} $-\delta t_M$};
    
            \node (dds) at ($(fpga.south) + (-0.5cm,0.2cm)$) [draw, rectangle, anchor=south] {DDS};
            \node (soc) at (fpga.south east) [draw, rectangle, anchor=south east, inner xsep=0.2cm, inner ysep =0.3cm] {\Large PS};
            \draw[->, thick, bronze] (soc) to[out=90, in=270] (del);
            \draw[->, thick, bronze] (soc.west |- dds.east)--(dds.east) node[below right, xshift=-0.05cm, yshift=0.05cm] {\small $f_{DDS}$};
    
            % \draw[->, thick, cadmiumgreen] (moku.north east) to[out=45, in=180] (clk104) node[xshift=-2.5cm, yshift=0.3cm] {Ext. Ref.};
            \draw[->, thick, magenta] (clk104) to [out=270, in=90] (adc.north) node[above left, xshift=0.3cm, yshift=0.2cm]{$-\delta t_D$};
            \draw[->, thick, cadmiumgreen] (dds) to[out=180, in=south] (dac);
            % \draw[->, thick, cadmiumgreen] (clk104) to[out=270, in=55] (dds);
        \end{tikzpicture}
        }
    \end{minipage}
    \begin{minipage}{0.5\linewidth}
    \centerline{\includegraphics[width=\textwidth]{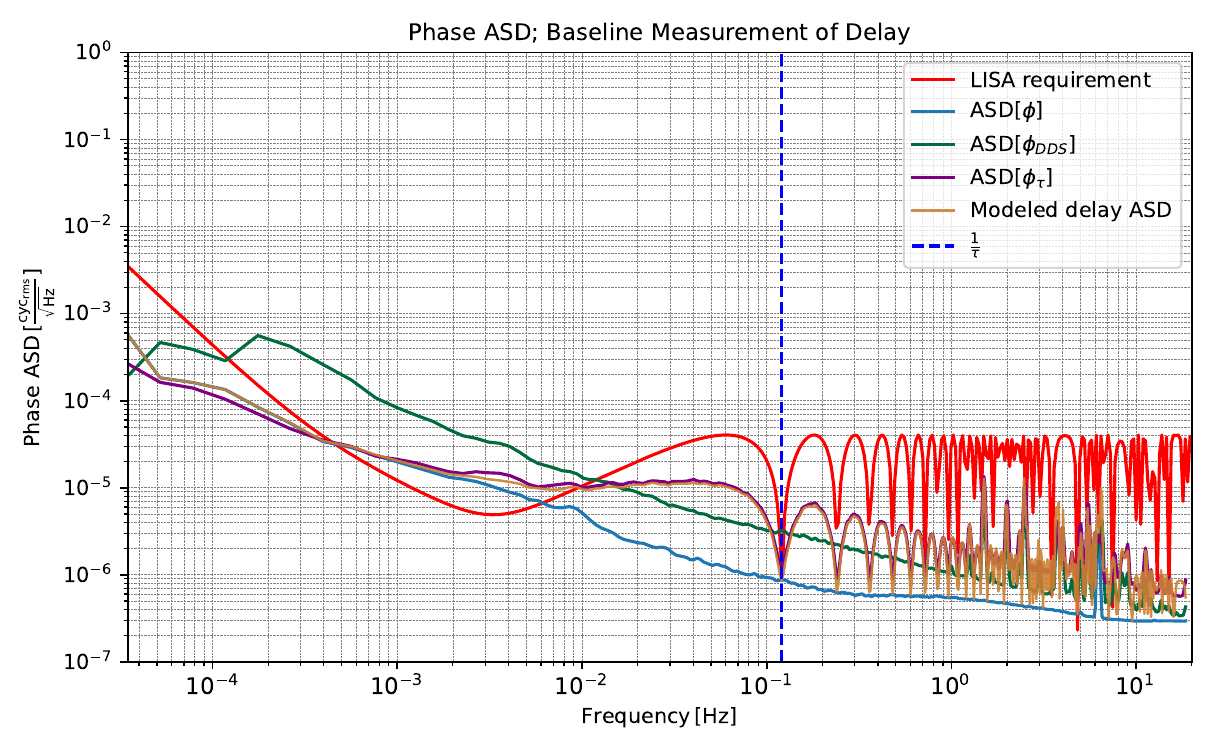}}
    \end{minipage}
    \caption{Left: Measurement setup for clock jitter analysis in the delay line, corresponding to Eq. \ref{eq:clocktrace}. The CLK104 clock board drives ADCs and DACs. Control signals are sent via Ethernet to the Processing System (PS), shown in the bottom-right, which feeds them to the FPGA. Right: plot of phase ASD. In blue, the ASD of the measured phase of the generated "carrier" signal, the green is that of the DDS signal (described in §\ref{subsec:acf}), the bronze is the model applied to them(Eq. \ref{eq:cjmodelASD}), and the purple is that of the delayed signal. The model fits well to the data. The LISA single-link phase noise requirement is shown in red.\protect\cite{babak}}

\end{figure}

Extending this logic to the spectral densities of these signals provides the plots shown in Figure \ref{fig:clockjitter}, where the model described in Equation \ref{eq:cjmodelASD} is applied using the spectral density estimations of the DDS tone and of the original generated signal. It shows good agreement with the measurement of the delayed signal.

This is commonly referred to as the \emph{delay comparison transfer function}, and frequently shows up in TDI calculations, where it is similar to how the full LISA constellation has no sensitivity at integer multiples of the light travel time. In this context however, it shows that the delay line does not add a significant amount of noise to the signal, and that this added noise can be accounted for and removed in post-processing. The delay line is therefore suitable for use in the hardware testbed.

% \begin{figure}
%     \centering
%     \includegraphics[width=0.6\textwidth]{Baseline_lasd.pdf}
%     \caption{Measurements corresponding to Figure \ref{fig:clockjitter} and Equaiton \ref{eq:cjmodel}. The blue line is the original signal, the purple line is the delayed signal, the orange line is the DDS signal, and the green line is the model applied using the original signal and the DDS signal. The model is a nearly perfect fit to the data.}
%     \label{fig:cjplot}
% \end{figure}

\subsection{TDI Test with MBHB Signal Injection}

As a demonstrative example, two delay lines were programmed into one board, each fed by the same signal source. These can be thought of as two separate optical paths that include a mirror reflecting each path back to the source --- the ``perfect transponder" model\cite{hartwig2021instrumental}. A separate signal generator with a higher frequency noise floor is used. The length of each ``arm" is set via control signals sent to the FPGA before the \texttt{enable} signal is raised. 

This reduces the expression given in Equation \ref{tdieq} with $\eta_{ij} + \mathbf{D}_{ij}\eta_{ji} \rightarrow \eta_{iji}$, giving the following TDI combination:
\begin{equation}
    X_1 = (1 - \mathcal{D}_{121})\eta_{131} - (1 - \mathcal{D}_{131})\eta_{121}.
\end{equation}

A time series of phase offsets corresponding to the strain of a Massive Black Hole Binary (MBHB) merger (calculated from the \texttt{IMRPhenomD} model set\cite{Khan_2016}) is loaded into the RAM of the on-chip processing system. This is used to feed these offsets into the FPGA so that they modulate the output of the delay line(s). These offsets can be scaled by an antenna response function; either an arbitrary constant value or one that evolves with time, calculated from orbital motion of the satellites. In the latter case, the response evolves hand-in-hand with Doppler modulation of the delayed signals and actuation of the delay times corresponding to relative satellite motion.

\begin{figure}
    \centering
    \begin{minipage}{0.49\linewidth}
        \centerline{\includegraphics[width=\textwidth]{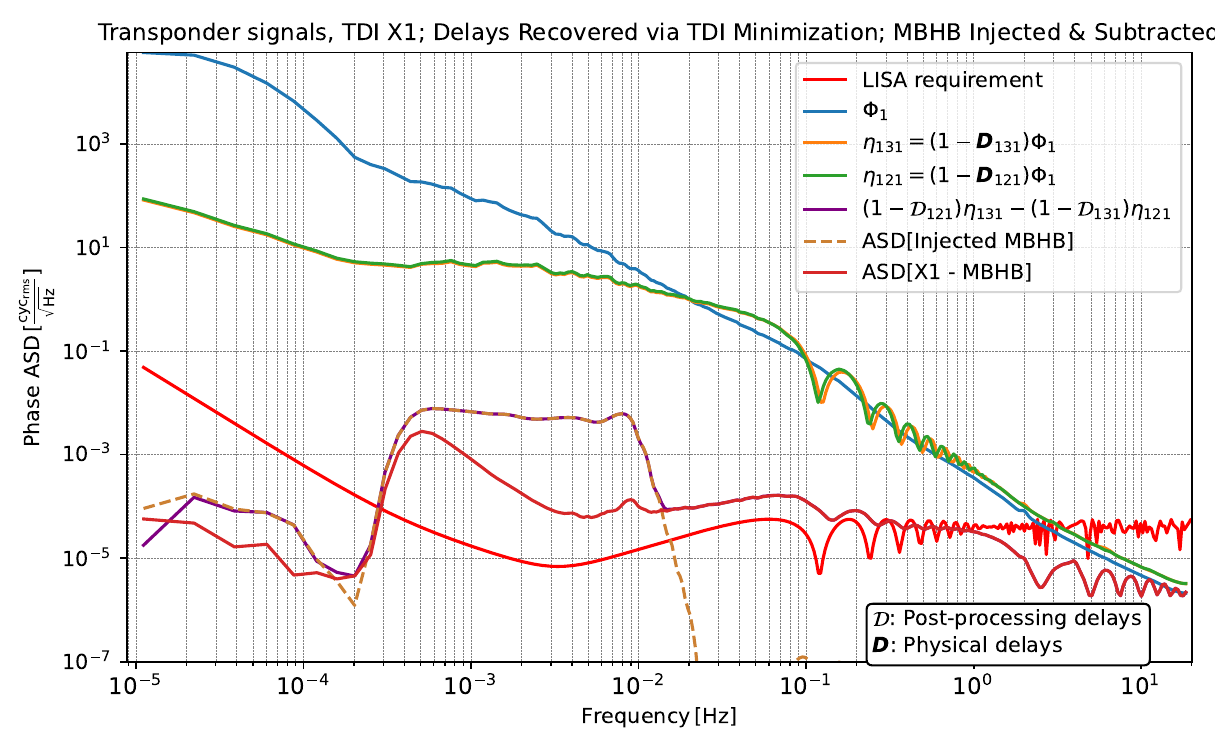}}
    \end{minipage}
    \begin{minipage}{0.49\linewidth}
        \centerline{\includegraphics[width=\textwidth]{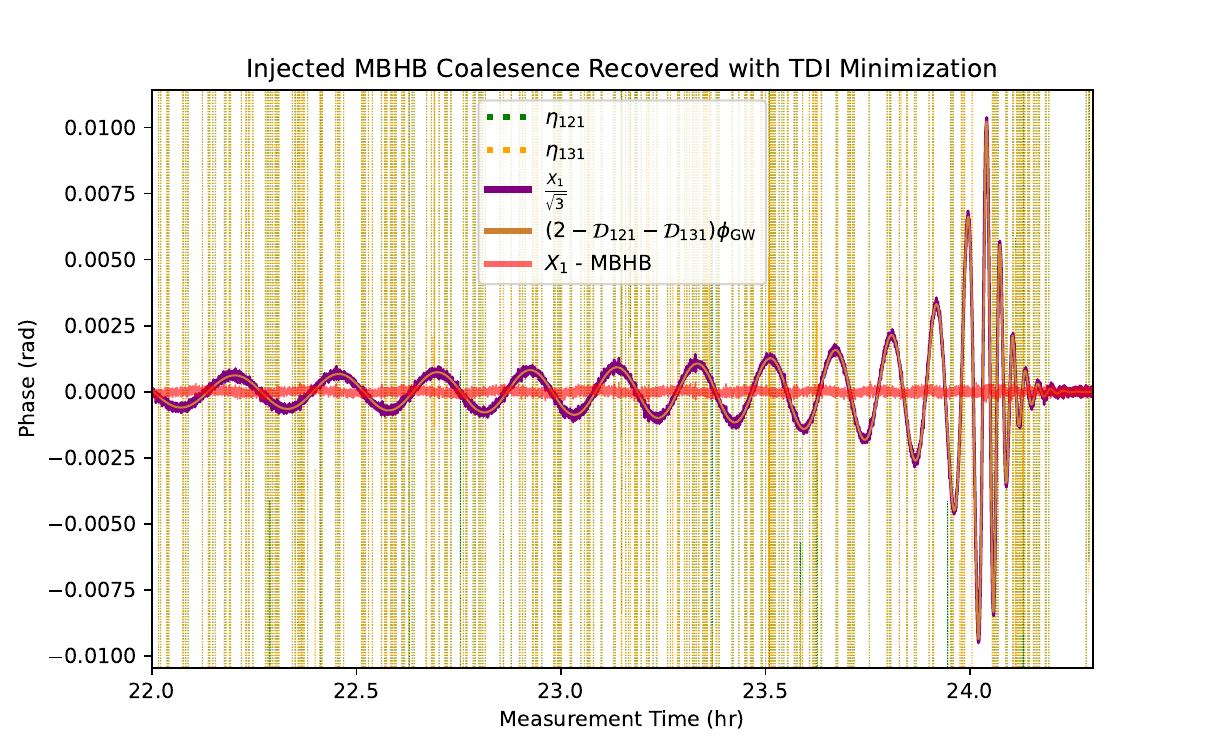}}
    \end{minipage}
    \caption{TDI combination (Equation \ref{tdieq}) suppresses the carrier frequency noise which would otherwise obscure the injected MBHB coalescence waveform, which was arbitrarily scaled by a $\frac{1}{\sqrt{3}}$ antenna response. The residual spectral noise is likely due to clocking difference in the clock which is used to update the injected offsets. This presents as phase dispersion between injection and measurement which can be seen for lower frequencies of the spectrum, whereas the amplitudes agree well individually.}
    \label{tdi}
\end{figure}

The undelayed and two delayed signal phases are measured. Applying the appropriate TDI combination removes the carrier frequency noise from the data as expected, revealing the characteristic form of a binary merger in the spectral density estimation and the injected waveform in the time series. The ASD[$X_1-\textnormal{MBHB}$] clearly does not reach the noise floor. This is likely due to timing differences between the clock driving data through the FPGA and the clock driving the PS feeding the offsets to the data. This should be resolved by buffering the offsets first in the data clock domain. It also suppresses the clock noise introduced by the delay line discussed in §\ref{subsec:acf}.\footnote{A satisfying process; everything cancels nicely to $X_1 = 0$ in the absence of an injected strain.} These are shown in Figure \ref{tdi}.

\section{Conclusions and Future Work}

In these proceedings, we have shown the first results of a new hardware-in-the-loop testbed that is capable of accurately simulating the inter-satellite delays in the LISA constellation, including the effects of timing jitter and Doppler frequency shifts. We have shown that the delay line does not add a prohibitive amount of noise to the signal, and that this added noise can be removed in post-processing. We have also shown that the TDI algorithms can be applied to the signals passed through this testbed, and that they successfully suppress the carrier frequency noise.  Additionally, this is the first delay testbed where realistic, complex GW strain can be injected and recovered.

Future tests and measurements will include simulations of orbital dynamics in the delay lines --- the frequencies of the carriers will change according to the pre-calculated relative motion of each spacecraft pairing, and the antenna response and delay times will be adjusted accordingly. The testbed will then be expanded to include simulations of all three ``spacecraft", such that independent clocking timebases are active in the system. This will allow for the testing of the clock timing transfer system, where the measurements done aboard each spacecraft are synchronized to each other via clock sideband beatnote measurements. The goal is to then combine all these factors with compounded strain signals, including confusion noise from many sources, to test the full LISA data analysis pipeline on data produced by LISA-like hardware.

\section*{References}
\bibliography{moriond}

\begin{thebibliography}{1}

\bibitem{Ghosh_2022}
Sourath Ghosh, Josep Sanjuan, and Guido Mueller.
\newblock Arm locking performance with the new {LISA} design.
\newblock {\em Classical and Quantum Gravity}, 39(11):115009, May 2022.

\bibitem{hartwig2021instrumental}
Olaf Hartwig.
\newblock {\em Instrumental modelling and noise reduction algorithms for the
  Laser Interferometer Space Antenna}.
\newblock PhD thesis, Leibniz University Hanover, Hanover, Germany, 2021.

\bibitem{babak}
S.~Babak et~al.
\newblock {LISA} sensitivity and {SNR} calculations.
\newblock 2021.

\bibitem{Khan_2016}
S.~Khan et~al.
\newblock Frequency-domain gravitational waves from nonprecessing black-hole
  binaries. {II. A} phenomenological model for the advanced detector era.
\newblock {\em Physical Review D}, 93(4), February 2016.

\end{thebibliography}

%%% manually generated bibliography
%\begin{thebibliography}{99}
%\bibitem{ja}C Jarlskog in {\em CP Violation}, ed. C Jarlskog
%(World Scientific, Singapore, 1988).
%\bibitem{ma}L. Maiani, \Journal{\PLB}{62}{183}{1976}.
%\bibitem{bu}J.D. Bjorken and I. Dunietz, \Journal{\PRD}{36}{2109}{1987}.
%\bibitem{bd}C.D. Buchanan {\it et al}, \Journal{\PRD}{45}{4088}{1992}.
%\end{thebibliography}

\end{document}